# A Stream-Suitable Kolmogorov-Smirnov-Type Test for Big Data Analysis

Hien D. Nguyen[1]

*Department of Mathematics and Statistics, La Trobe University, Bundoora 3086, Australia*

**Abstract**

Big Data has become an ever more commonplace setting that is encountered by data analysts. In the Big Data setting, analysts are faced with very large numbers of observations as well as data that arrive as a stream, both of which are phenomena that many traditional statistical techniques are unable to contend with. Unfortunately, many of these traditional techniques are useful and cannot be discarded. One such technique is the Kolmogorov-Smirnov (KS) test for goodness-of-fit (GoF). A Big Data and stream-appropriate KS-type test is derived via the chunked-and-averaged (CA) estimator paradigm. The new test is termed the CAKS GoF test. The CAKS test statistic is proved to be asymptotically normal, allowing for the large sample testing of GoF. Furthermore, theoretical results demonstrate that the CAKS test is consistent against both fixed alternatives, where the null and the true data generating distribution are a fixed distance apart, and alternatives that approach the null at a slow enough rate. Numerical results demonstrate that the CAKS test is effective in identifying deviation in the distribution with respect to changes in mean, variance, and shape. Furthermore, it is found that the CAKS test is faster than the KS test, for large numbers of observation, and can be applied to sample sizes of $10^9$ and beyond.

[1]h.nguyen5@latrobe.edu.au

## 1. Introduction

Big Data has become an ever more pervasive setting in which the modern data analyst must operate. The American National Institute of Standards and Technology (NIST) considers Big Data to be '... data of which the data volume, acquisition speed, or data representation limits the capacity of using traditional relational methods to conduct effective analysis ...' (Chen et al., 2014) and Jacobs (2009) defines Big Data as '... data whose size forces us to look beyond the tried-and-true methods that are prevalent at that time'.

Unfortunately, there are some tried-and-true techniques that are fundamental to the practice of data analysis, which cannot be discarded, and which must be adapted and modified to overcome the challenges presented by the increasing computational demands of the Big Data setting. One among these fundamental techniques is the Kolmogorov-Smirnov (KS) goodness-of-fit (GoF) test (Kolmogorov, 1933; Smirnov, 1948), for assessing the fit of a given data set to a known probability distribution model. For instance, it is suggested by Buoncristiano et al. (2015) that the development of a fast and Big Data-appropriate implementation of the KS test is a key research challenge in modern exploratory data analysis.

Outside of exploratory data analysis, there are many other applications for the KS test that are within the domain of Big Data. These include conducting gene set enrichment analysis in systems biology (Clark and Ma'ayan, 2011); detecting workload changes in database management Abad et al. (2012); assessing the correctness of multiple imputation algorithms Nguyen et al. (2013) for large data sets; change detection in analysis of streamed data Tran et al. (2014);

assessing the distribution of chemicals in large geological samples (Vermeesch and Garzaanti, 2015); and testing of random numbers from random numbers for use in cryptographic use (see e.g. Demirhan and Bitirim, 2016, and references therein). A further application of the KS test is for assessing chain stability in MCMC (Markov chain Monte Carlo) simulations for Bayesian analysis (Gruet et al., 1998; Robert and Casella, 2010, Ch. 8). Here, the availability of KS tests for very large samples is particularly important as modern Bayesian analyses can feature MCMC chains that can be more than $10^7$ observations long; see for example Drummond et al. (2006) and Macqueen et al. (2014). More Big Data applications for the KS test are cited by Lall (2015), including analyses of astrological, wireless sensor networks, internet measurements data.

In this article, we consider an algorithm for GoF testing that is based on the KS statistic, which is appropriate for online testing in Big Data settings. That is, our test is appropriate for use when the data arrive as a stream rather than in a batch, as is the nature of the offline setting. As noted by Bifet et al. (2010), a successful algorithm in the online setting must satisfy three requirements. Namely, (i) the algorithm must be able to inspect the examples in a small number at a time, and inspect each example at most once; (ii) the algorithm must be able to operate with a limited amount of memory; and (iii) the algorithm must be able to work in a limited amount of time, especially when compared to its batch counterpart. We shall demonstrate that our GoF testing algorithm satisfies all of the requirements of Bifet et al. (2010).

The presented algorithm is based on the online learning approach of Nguyen (2017), which is in turn based on the work on chunked-and-averaged (CA) estimators of Li et al. (2013), Matloff (2016a, Ch. 13), and Matloff (2016b); see also Nguyen and McLachlan (2017). As such, we call our testing procedure the chunked-and-averaged Kolmogorov-Smirnov test (CAKS). Along with the derivation, description, and presentation of the CAKS GoF test, we also present some theoretical results regarding the asymptotic performance of the procedure. That is, we prove that the CAKS test, like the KS test that it is based on, is asymptotically point-wise consistent against fixed alternatives as well as uniformly consistent against alternatives that approach the null at a slow enough rate (cf. Lehmann and Romano, 2005, Thms. 14.2.1 and 14.2.2). A numerical study is also conducted to demonstrate the use of the CAKS test as well as its computational advantage over its batch counterpart.

We note that our work is parallel to that of Lall (2015), who also considers the problem of performing KS GoF tests in the streamed context. Our work differs to that of Lall (2015) in a number of ways. Firstly, we consider the construction of a test statistic that is based on the KS test statistic, and which has a known asymptotic distribution, whereas Lall (2015) considers the error-bounded approximation of the KS test statistic via the use of quantile sketches (see e.g. Wang et al., 2013, and references therein) and to be used with the usual KS sampling distribution. Secondly, our method is designed to allow for streamed testing of very large data sets of sizes of $10^9$ observations, and beyond, whereas Lall (2015) only considers data of sizes up to $10^5$ observations. Thirdly, our algorithm requires a fixed amount of memory, regardless of the final length $N$ of the data stream, whereas the approach from Lall (2015) requires storage of $O\left(N^{1/2}\log N\right)$ samples. We do note, however, that Lall (2015) also presents a method for conducting streamed two-sample KS tests, whereas we only concentrate on the one sample setting.

The article proceeds as follows. In Section 2, we derive the CAKS GoF testing algorithm from the framework of CA estimators. In Section 3, theoretical properties of the CAKS test are proved. In Section 4, a numerical study is conducted to assess the statistical and computation performance of the CAKS testing algorithm. Finally, conclusions are drawn in Section 5.

## 2. The CAKS GoF Test

### *2.1. Chunked-and-Averaged Estimators*

Let $\mathbf{X} = \{X_i\}_{i=1}^N$ be an IID (independent and identically distributed) random sample from some probability distribution model $F_0$, with unknown parameter $\theta = \theta(F_0)$. Here, $\theta \in \mathbb{R}$ and $X_i \in \mathbb{R}$, where $i \in [N]$ and $[N] = \{1, \ldots, N\}$. Suppose that we can arrange $\mathbf{X}$ in the $T \times J$ array

$$\begin{array}{cccc} X_{11}, & X_{12}, & \ldots, & X_{1J}, \\ X_{21}, & X_{22}, & \ldots, & X_{2J}, \\ \ldots, & \ldots, & \ldots, & \ldots, \\ X_{T1}, & X_{T2}, & \ldots, & X_{TJ}, \end{array} \tag{1}$$

where $X_{tj} = X_{(t-1)J+j}$ ($t \in [T]$, $j \in [J]$) can be used to map from (1) back to $\mathbf{X}$, and $N = T \times J$. We call each row of (1) a chunk of the data and denote the subsample belonging to each chunk as $\mathbf{X}_t = \{X_{tj}\}_{j=1}^J$. The choice of subscript $t$ is used here to indicate time. That is, we can interpret the chunks as being a stream of data whereupon we receive a constant stream of $J$ observations (i.e. $\mathbf{X}_t$) at each time period $t$ until time period $T$, whereupon all $N$ observations will have been obtained.

Suppose that $\theta$ can be estimated via some batch estimator $\hat{\Theta} = \hat{\theta}(\mathbf{X})$ , where $\hat{\theta}$ takes an arbitrary number of data points as an input. We define the same estimator on each of the chunk as the 'chunked' estimator, and write it as $\hat{\Theta}_t = \hat{\theta}(\mathbf{X}_t)$. Assuming that they exist, we also write the mean and variance of the chunked estimator for chunks of size $J$ as $\mu_J = \mathbb{E}\left(\hat{\Theta}_t\right)$ and $\sigma_J^2 = \mathrm{Var}\left(\hat{\Theta}_t\right)$, respectively. We finally define the CA estimator as

$$\bar{\Theta}_T = T^{-1} \sum_{t=1}^{T} \hat{\Theta}_t. \tag{2}$$

Given a stream of realizations of $\mathbf{X}$, which we shall write as $\mathbf{x} = \{x_i\}_{i=1}^{N}$, and its corresponding chunks $\mathbf{x}_t = \{x_{tj}\}_{j=1}^{J}$, we can write the realized chunked estimates as $\hat{\theta}_t = \hat{\theta}(\mathbf{x}_t)$. The CA estimate based on the chunks $\mathbf{x}_t$ arriving in a stream, can be computed in an online manner via the iteration scheme:

$$\bar{\theta}_t = \frac{(t-1)\,\bar{\theta}_{t-1} + \hat{\theta}_t}{t}, \tag{3}$$

for $t \in [T]$, where $\bar{\theta}_0$ can be set arbitrarily. We note that scheme (3) implies that at any time $t$, we only require storage of the $J$ observations in $\mathbf{x}_t$ and the single previous iterate $\bar{\theta}_{t-1}$. We further, observe that (3) produces the estimated value for (2), when $t = T$, given the realization $\mathbf{x}$.

Denote non-stochastic convergence and convergence in distribution by $\rightarrow$ and $\rightsquigarrow$, respectively. Further, let $Z$ denote a standard normal random variable, with probability distribution function $\Phi$. The asymptotic normality of (2) can be obtained via the following result of Li et al. (2013); see also Nguyen and McLachlan (2017).

**Lemma 1.** *Assume that* $\mathbf{X}$ *is an IID sample from some probability distribution model* $F_0$*. If* $\mu_J$ *and* $\sigma_J^2$ *exist, then as* $T \rightarrow \infty$*,* $T^{1/2}\sigma_J^{-1}\left(\bar{\Theta}_T - \mu_J\right) \rightsquigarrow Z$*.*

*Remark* 1. We note that Lemma 1 implies that if $\mu_J \neq \theta$, then $\bar{\Theta}_T$ will be a biased and inconsistent estimator of the parameter of interest. Although this is a problem in general, as correct estimation of $\theta$ is usual the concern, here we require the asymptotic normality property that is obtained from Lemma 1, and nothing further.

*2.2. The Kolmogorov-Smirnov Statistic*

The KS test centers around the use of the Kolmogorov metric between two probability distributions $F_1$ and $F_2$ over a common domain $\mathbb{X}$. We can write the Kolmogorov metric between $F_1$ and $F_2$ as

$$d\left(F_1, F_2\right) = \sup_{x \in \mathbb{X}} \left|F_1\left(x\right) - F_2\left(x\right)\right|. \tag{4}$$

Here, (4) is a proper metric function under the usual definition (cf. DasGupta, 2011, Def. 15.1).

As in Section 2.1, let $\mathbf{X}$ be a random sample that arises from some probability distribution model $F_0$. Suppose that we wish to test the null hypothesis

$$H_0 : F_0 = F, \tag{5}$$

against the alternative

$$H_1 : F_0 \neq F, \tag{6}$$

for some probability distribution $F$. Based on the random sample, the KS test statistic is defined as

$$K\left(\mathbf{X}\right) = d\left(\hat{F}_N, F\right), \tag{7}$$

where

$$\hat{F}_N\left(x\right) = N^{-1} \sum_{i=1}^{N} \mathbb{I}\left(X_i \leq x\right)$$

is the empirical distribution function based on the sample $\mathbf{X}$, and $\mathbb{I}\left(A\right)$ is the indicator function that returns 1 if $A$ is true and 0 otherwise.

We can observe that the null distribution 7, for any fixed $N$, is invariant to $F$ and thus the analysis of the behavior of 7 under $H_0$ can be significantly simplified by analyzing the case where $F$ is the uniform distribution over the unit interval. Under a uniform null, (7) can be written as:

$$K(\mathbf{X}) = \max_{i \in [N]} \max \left\{ X_{(i)} - \frac{i-1}{N}, \frac{i}{N} - X_{(i)} \right\}, \quad (8)$$

where $X_{(i)}$ is the $i$th order statistic of $\mathbf{X}$ (cf. Wang et al., 2003). Using form (8) and the recursive techniques from Durbin (1972, Sec. 2.4), Wang et al. (2003) determined that the probability distribution of the KS test statistic, for any fixed $N$, is a piecewise polynomial function. This implies that the distribution function of (7) is measurable for any fixed $N$, and therefore since $K \in [0,1]$ by definition, (7) has all of its moments.

An algorithm for generating the distribution of (7) for any $N$ is provided in Wang et al. (2003), for the computation of KS test probabilities. Unfortunately, due to the highly-recursive nature of the algorithm, it is only suitable for rapid computation of probabilities when $N$ is small (i.e. $N \le 100$). When $N$ is large, the asymptotic result:

$$\mathbb{P}\left(N^{1/2} K(\mathbf{X}) > k\right) \to 2 \sum_{j=1}^{\infty} (-1)^{j+1} \exp\left(-2j^2k^2\right), \quad (9)$$

for $k > 0$, is generally used to draw inference from (7) (cf. Lehmann and Romano, 2005, Sec. 14.2.1). It is noted in Wang et al. (2003) that (9) converges rather slowly to the true distribution, even for traditionally large numbers of observations (e.g. up to $N = 4096$).

### *2.3. Construction of the CAKS Test*

Consider, as before, that $\mathbf{X}$ arises from some distribution $F_0$. If we wish to estimate the parameter $\theta = d(F_0, F)$, for some interesting $F$, then we can utilize the KS test statistic (7). That is, $\hat{\Theta} = K(\mathbf{X})$ is a batch statistic that estimates $\theta$. Given the chunks $\mathbf{X}_t$, the CA estimator that is based on (7) has form (2), where $\hat{\Theta}_t = K(\mathbf{X}_j) = d\left(\hat{F}_{tJ}, F\right)$ and

$$\hat{F}_{tJ}(\mathbf{X}_J) = J^{-1}\sum_{j=1}^{J}\mathbb{I}(X_{tj} \leq x).$$

We name this CA estimator the CAKS statistic.

Based on the fact that the distribution of (7) is piecewise polynomial, and hence measurable, for any $N$, we obtain the existence of $\mu_J$ and $\sigma_J^2$, for each $J$. Thus, via Lemma 1, we have the following asymptotic normality result for the CAKS statistic.

**Proposition 1.** *For any distribution function $F$, let*

$$\bar{\Theta}_T = T^{-1}\sum_{t=1}^{T} d\left(\hat{F}_{tJ}, F\right) \tag{10}$$

*be the CAKS statistic. If $\mathbf{X}$ is an IID sample from $F_0$, then $T^{1/2}\sigma_J^{-1}\left(\bar{\Theta}_T - \mu_J\right) \rightsquigarrow Z$, where $\mu_J = \mathbb{E}\left[d\left(\hat{F}_{tJ}, F\right)\right]$ and $\sigma_J^2 = Var\left[d\left(\hat{F}_{tJ}, F\right)\right]$.*

Given Proposition 1 and some predetermined test size $\alpha$, we can construct a test of (5) against (6) via the usual $Z$-test rule:

$$r_\alpha(\mathbf{X}) = \begin{cases} 0, & \text{if } 1 - \Phi\left(T^{1/2}\sigma_J^{-1}\left[\bar{\Theta}_T - \mu_J\right]\right) > \alpha, \\ 1, & \text{otherwise,} \end{cases} \tag{11}$$

Here, $r_\alpha(\mathbf{X}) = 1$ indicates a rejection of the null and 0 otherwise. By virtue of the size and by the asymptotic normality result, $\mathbb{P}(r_\alpha(\mathbf{X}) = 1 | F_0 = F) \to \alpha$ as $T \to \infty$. There is no blowup in the asymptotic size of the test as is guaranteed by the existence of all of the moments of (7), and via Lehmann and Romano (2005, Lem. 11.4.1).

*Remark* 2. As mentioned in Remark 1, the CAKS statistic $\bar{\Theta}_T$ is only a consis-tent estimator of $\mu_J$, and not $\theta$. Furthermore, both $\mu_J$ and $\sigma_J^2$ are specific to each choice of $J$. As such, we have provided a table of well-estimated values for $J^{1/2} \times \mu_J$ and $J \times \sigma_J^2$, which are each obtained from $10^6$ Monte Carlo replicates. The results are presented in Table 1. The choice to premultiply $\sigma_J$ and $\sigma_J^2$ by $J^{1/2}$ and $J$, respectively, is to make their values comparable to the asymptotic mean and variances that are obtainable from (9). The asymptotic mean and variance of $N^{1/2}K(\mathbf{X})$, as $N \to \infty$, are $(\pi/2)^{1/2} \log 2$ and $\pi^2/12 - (\pi/2)^{1/2} \log 2$, respectively (cf. Wang et al., 2003), and are given in decimal form on the last line of Table 1. We note that the notation $a\text{E}b$ is short for $a \times 10^b$.

Table 1: Values for $J^{1/2} \times \mu_J$ and $J \times \sigma_J^2$, estimated via $10^6$ Monte Carlo replicates.

| $J$ | $J^{1/2} \times \mu_J$ | $J \times \sigma_J^2$ |
|---|---|---|
| 1.00E+02 | 0.8525199 | 0.06734524 |
| 2.00E+02 | 0.8567999 | 0.06759134 |
| 5.00E+02 | 0.8613619 | 0.06758775 |
| 1.00E+03 | 0.8638072 | 0.06768931 |
| 2.00E+03 | 0.8649057 | 0.06747976 |
| 5.00E+03 | 0.8661168 | 0.06769269 |
| 1.00E+04 | 0.8670493 | 0.06772049 |
| 2.00E+04 | 0.8674296 | 0.06777601 |
| 5.00E+04 | 0.8679541 | 0.06782049 |
| 1.00E+05 | 0.8683573 | 0.06788714 |
| 2.00E+05 | 0.8684477 | 0.06780899 |
| 5.00E+05 | 0.8685257 | 0.06765810 |
| 1.00E+06 | 0.8685212 | 0.06787872 |
| $J \to \infty$ | 0.8687312 | 0.06777320 |

## 3. Theoretical Results

The following pair of results show that rule (11) is both point-wise consistent in power, and consistent under a sequence of alternatives that converge to the null. The notations $a.s.$ and $\overset{a.s.}{\to}$ stand for almost surely, and almost sure convergence, respectively.

**Proposition 2.** *If* $\mathbf{X}$ *is an IID sample from some distribution* $F_1 \neq F_0 = F$ *such that* $\mathbb{E}\left[d\left(\hat{F}_{tJ}F\right)\right] \geq \delta > \mu_J$*, then as* $T \to \infty$*,* $r_\alpha(\mathbf{X}) = 1$*, a.s., for any* $\alpha \in (0,1)$ *and* $J \in \mathbb{N}$*.*

*Proof.* Under the hypothesis of the proposition and by definition, we obtain the inequality $\bar{\Theta}_T \geq \delta > \mu_J$, a.s. for sufficiently large $T$, by the law of large numbers. Since $\Phi$ is an increasing function, we have

$$1 - \Phi\left(\frac{\bar{\Theta}_T - \mu_J}{T^{-1/2}\sigma_J}\right) \leq 1 - \Phi\left(T^{1/2}\frac{\delta - \mu_J}{\sigma_J}\right)$$

a.s. by continuous mapping, where the right hand side converges to $1 - 1 = 0$ as $T \to \infty$, since $\delta > \mu_J$. We obtain the desired result via definition (11). □

**Proposition 3.** *If* $\mathbf{X}$ *is an IID sample from* $F_T \neq F_0 = F$ *such that* $\mathbb{E}\left[d\left(\hat{F}_{tJ}, F\right)\right] \geq T^{-1/2}\delta_T + \mu_J$ *for all* $T \in \mathbb{N}$*, where* $\delta_T \to \infty$*, then as* $T \to \infty$*,* $r_\alpha(\mathbf{X}) = 1$*, a.s., for any* $\alpha \in (0,1)$ *and* $J \in \mathbb{N}$*.*

*Proof.* Under the hypothesis of the proposition and by definition, we obtain the inequality $\bar{\Theta}_T - \mu_J \geq T^{-1/2}\delta_T$, a.s. for sufficiently large $T$, by the law of large numbers. Since $\Phi$ is an increasing function, we have

$$\begin{aligned} 1 - \Phi\left(\frac{\bar{\Theta}_T - \mu_J}{T^{-1/2}\sigma_J}\right) &\leq 1 - \Phi\left(T^{1/2}\frac{T^{-1/2}\delta_T}{\sigma_J}\right) \\ &= 1 - \Phi\left(\frac{\delta_T}{\sigma_J}\right) \end{aligned}$$

a.s. by continuous mapping, where the right hand side converges to $1 - 1 = 0$ as $T \to \infty$, since $\delta_T \to \infty$. We obtain the desired result via definition (11). □

*Remark* 3. Propositions 2 and 3 mirror the consistency results regarding the usual KS test, as presented in Lehmann and Romano (2005, Thms. 14.2.1 and 14.2.2). Together, they allow practitioners to be confident of the power of the CAKS test, even when the Kolmogorov metric under the alternative becomes arbitrarily close to the null value as more observations are obtained.

## 4. Numerical Study

We conduct a numerical study via a set of three simulation scenarios S1–S3. All computations conducted for the study are performed on a MacBook Pro with a 2.2 GHz Intel Core i7 processor, 16 GB of 1600 MHz DDR3 memory, and a 500 GB SSD. Furthermore, implementations of all computational processes and algorithms are performed via the R programming language and interpreter environment (R Core Team, 2016).

In all three simulation scenarios, S1–S3, the same basic setup is used. That is, $T \in \{100, 1000\}$ chunks of observations of sizes $J \in \bigcup_{i=2}^{5} \{1, 2, 5\} \times 10^i \cup \left\{10^6\right\}$ are simulated from a distribution that depends on the simulation scenario. The CAKS test, using the $N = T \times J$, is conducted for the null hypothesis $H_0 : F_0 = \Phi$ . In scenario S1, we simulate data from a normal distribution with variance 1 but varying means $\mu \in \{0, 0.001, 0.01, 0.1, 1\}$. In scenario S2, we simulate data from a normal distribution with mean 0 but with varying variances $\sigma^2 \in \left\{0.8^2, 0.9^2, 0.95^2, 1.05^2, 1.1^2, 1.2^2\right\}$. In scenario S3, we simulate data from a $t$ distribution with degrees of freedom $\text{df} \in \{1, 5, 10, 50, 100\}$.

For each of the scenarios and at each of the variable values, we replicate the CAKS test 100 times. From the replicates, we note the total number of rejections at the size $\alpha = 0.1$ as well as the average time taken to conduct the tests. The results of the numerical study are reported in tables 2–7. The average times are given in seconds and timing was conducted using the proc.time() function.

### *4.1. Results*

Upon inspection of Tables 2–7, we can make the following interpretations. Firstly, the first column of Table 2, corresponding to the $\mu = 0$ case of S1,

Table 2: CAKS test results for the cases $\mu \in \{0, 0.001, 0.01\}$ of scenario S1.

| $\mu$ | $T$ | $J$ | Rejections | Avg. Time | $\mu$ | $T$ | $J$ | Rejections | Avg. Time | $\mu$ | $T$ | $J$ | Rejections | Avg. Time |
|---|---|---|---|---|---|---|---|---|---|---|---|---|---|---|
| 0 | 100 | 1.00E+02 | 4 | 4.52E-03 | 0.001 | 100 | 1.00E+02 | 10 | 4.84E-03 | 0.01 | 100 | 1.00E+02 | 10 | 4.95E-03 |
| 0 | 100 | 2.00E+02 | 11 | 6.53E-03 | 0.001 | 100 | 2.00E+02 | 10 | 7.02E-03 | 0.01 | 100 | 2.00E+02 | 9 | 7.10E-03 |
| 0 | 100 | 5.00E+02 | 6 | 1.30E-02 | 0.001 | 100 | 5.00E+02 | 15 | 1.36E-02 | 0.01 | 100 | 5.00E+02 | 14 | 1.40E-02 |
| 0 | 100 | 1.00E+03 | 12 | 2.42E-02 | 0.001 | 100 | 1.00E+03 | 13 | 2.46E-02 | 0.01 | 100 | 1.00E+03 | 22 | 2.54E-02 |
| 0 | 100 | 2.00E+03 | 11 | 4.72E-02 | 0.001 | 100 | 2.00E+03 | 14 | 4.77E-02 | 0.01 | 100 | 2.00E+03 | 40 | 4.91E-02 |
| 0 | 100 | 5.00E+03 | 16 | 1.19E-01 | 0.001 | 100 | 5.00E+03 | 15 | 1.18E-01 | 0.01 | 100 | 5.00E+03 | 89 | 1.22E-01 |
| 0 | 100 | 1.00E+04 | 10 | 2.42E-01 | 0.001 | 100 | 1.00E+04 | 16 | 2.40E-01 | 0.01 | 100 | 1.00E+04 | 99 | 2.47E-01 |
| 0 | 100 | 2.00E+04 | 9 | 4.98E-01 | 0.001 | 100 | 2.00E+04 | 16 | 4.92E-01 | 0.01 | 100 | 2.00E+04 | 100 | 5.08E-01 |
| 0 | 100 | 5.00E+04 | 6 | 1.32E+00 | 0.001 | 100 | 5.00E+04 | 12 | 1.30E+00 | 0.01 | 100 | 5.00E+04 | 100 | 1.34E+00 |
| 0 | 100 | 1.00E+05 | 4 | 2.79E+00 | 0.001 | 100 | 1.00E+05 | 23 | 2.83E+00 | 0.01 | 100 | 1.00E+05 | 100 | 2.92E+00 |
| 0 | 100 | 2.00E+05 | 10 | 5.75E+00 | 0.001 | 100 | 2.00E+05 | 36 | 5.90E+00 | 0.01 | 100 | 2.00E+05 | 100 | 5.84E+00 |
| 0 | 100 | 5.00E+05 | 11 | 1.44E+01 | 0.001 | 100 | 5.00E+05 | 79 | 1.44E+01 | 0.01 | 100 | 5.00E+05 | 100 | 1.48E+01 |
| 0 | 100 | 1.00E+06 | 10 | 2.91E+01 | 0.001 | 100 | 1.00E+06 | 100 | 2.92E+01 | 0.01 | 100 | 1.00E+06 | 100 | 3.01E+01 |
| 0 | 1000 | 1.00E+02 | 8 | 4.68E-02 | 0.001 | 1000 | 1.00E+02 | 7 | 4.76E-02 | 0.01 | 1000 | 1.00E+02 | 9 | 4.85E-02 |
| 0 | 1000 | 2.00E+02 | 10 | 6.80E-02 | 0.001 | 1000 | 2.00E+02 | 15 | 6.88E-02 | 0.01 | 1000 | 2.00E+02 | 19 | 7.06E-02 |
| 0 | 1000 | 5.00E+02 | 8 | 1.34E-01 | 0.001 | 1000 | 5.00E+02 | 8 | 1.34E-01 | 0.01 | 1000 | 5.00E+02 | 31 | 1.38E-01 |
| 0 | 1000 | 1.00E+03 | 9 | 2.45E-01 | 0.001 | 1000 | 1.00E+03 | 4 | 2.45E-01 | 0.01 | 1000 | 1.00E+03 | 56 | 2.53E-01 |
| 0 | 1000 | 2.00E+03 | 8 | 4.75E-01 | 0.001 | 1000 | 2.00E+03 | 12 | 4.78E-01 | 0.01 | 1000 | 2.00E+03 | 98 | 4.87E-01 |
| 0 | 1000 | 5.00E+03 | 13 | 1.19E+00 | 0.001 | 1000 | 5.00E+03 | 15 | 1.19E+00 | 0.01 | 1000 | 5.00E+03 | 100 | 1.20E+00 |
| 0 | 1000 | 1.00E+04 | 11 | 2.42E+00 | 0.001 | 1000 | 1.00E+04 | 10 | 2.42E+00 | 0.01 | 1000 | 1.00E+04 | 100 | 2.43E+00 |
| 0 | 1000 | 2.00E+04 | 9 | 4.95E+00 | 0.001 | 1000 | 2.00E+04 | 19 | 4.97E+00 | 0.01 | 1000 | 2.00E+04 | 100 | 4.99E+00 |
| 0 | 1000 | 5.00E+04 | 4 | 1.33E+01 | 0.001 | 1000 | 5.00E+04 | 29 | 1.33E+01 | 0.01 | 1000 | 5.00E+04 | 100 | 1.34E+01 |
| 0 | 1000 | 1.00E+05 | 14 | 2.82E+01 | 0.001 | 1000 | 1.00E+05 | 65 | 2.82E+01 | 0.01 | 1000 | 1.00E+05 | 100 | 2.95E+01 |
| 0 | 1000 | 2.00E+05 | 5 | 5.83E+01 | 0.001 | 1000 | 2.00E+05 | 93 | 5.84E+01 | 0.01 | 1000 | 2.00E+05 | 100 | 5.89E+01 |
| 0 | 1000 | 5.00E+05 | 9 | 1.43E+02 | 0.001 | 1000 | 5.00E+05 | 100 | 1.48E+02 | 0.01 | 1000 | 5.00E+05 | 100 | 1.48E+02 |
| 0 | 1000 | 1.00E+06 | 16 | 2.95E+02 | 0.001 | 1000 | 1.00E+06 | 100 | 3.00E+02 | 0.01 | 1000 | 1.00E+06 | 100 | 3.02E+02 |

Table 3: CAKS test results for the cases $\mu \in \{0.1, 1\}$ of scenario S1.

| $\mu$ | $T$ | $J$ | Rejections | Avg. Time | $\mu$ | $T$ | $J$ | Rejections | Avg. Time |
|---|---|---|---|---|---|---|---|---|---|
| 0.1 | 100 | 1.00E+02 | 100 | 5.27E-02 | 1 | 100 | 1.00E+02 | 100 | 4.87E-02 |
| 0.1 | 100 | 2.00E+02 | 100 | 7.53E-02 | 1 | 100 | 2.00E+02 | 100 | 6.90E-02 |
| 0.1 | 100 | 5.00E+02 | 100 | 1.44E-01 | 1 | 100 | 5.00E+02 | 100 | 1.34E-01 |
| 0.1 | 100 | 1.00E+03 | 100 | 2.66E-01 | 1 | 100 | 1.00E+03 | 100 | 2.45E-01 |
| 0.1 | 100 | 2.00E+03 | 100 | 5.14E-01 | 1 | 100 | 2.00E+03 | 100 | 4.72E-01 |
| 0.1 | 100 | 5.00E+03 | 100 | 1.30E+00 | 1 | 100 | 5.00E+03 | 100 | 1.18E+00 |
| 0.1 | 100 | 1.00E+04 | 100 | 2.65E+00 | 1 | 100 | 1.00E+04 | 100 | 2.42E+00 |
| 0.1 | 100 | 2.00E+04 | 100 | 5.46E+00 | 1 | 100 | 2.00E+04 | 100 | 4.98E+00 |
| 0.1 | 100 | 5.00E+04 | 100 | 1.46E+01 | 1 | 100 | 5.00E+04 | 100 | 1.31E+01 |
| 0.1 | 100 | 1.00E+05 | 100 | 3.08E+01 | 1 | 100 | 1.00E+05 | 100 | 2.91E+01 |
| 0.1 | 100 | 2.00E+05 | 100 | 5.91E+00 | 1 | 100 | 2.00E+05 | 100 | 5.77E+00 |
| 0.1 | 100 | 5.00E+05 | 100 | 1.53E+02 | 1 | 100 | 5.00E+05 | 100 | 1.46E+02 |
| 0.1 | 100 | 1.00E+06 | 100 | 3.09E+02 | 1 | 100 | 1.00E+06 | 100 | 3.07E+02 |
| 0.1 | 1000 | 1.00E+02 | 100 | 5.07E-01 | 1 | 1000 | 1.00E+02 | 100 | 4.79E-01 |
| 0.1 | 1000 | 2.00E+02 | 100 | 7.34E-01 | 1 | 1000 | 2.00E+02 | 100 | 6.92E-01 |
| 0.1 | 1000 | 5.00E+02 | 100 | 1.44E+00 | 1 | 1000 | 5.00E+02 | 100 | 1.35E+00 |
| 0.1 | 1000 | 1.00E+03 | 100 | 2.59E+00 | 1 | 1000 | 1.00E+03 | 100 | 2.45E+00 |
| 0.1 | 1000 | 2.00E+03 | 100 | 5.05E+00 | 1 | 1000 | 2.00E+03 | 100 | 4.76E+00 |
| 0.1 | 1000 | 5.00E+03 | 100 | 1.28E+01 | 1 | 1000 | 5.00E+03 | 100 | 1.20E+01 |
| 0.1 | 1000 | 1.00E+04 | 100 | 2.62E+01 | 1 | 1000 | 1.00E+04 | 100 | 2.55E+01 |
| 0.1 | 1000 | 2.00E+04 | 100 | 5.32E+01 | 1 | 1000 | 2.00E+04 | 100 | 5.12E+01 |
| 0.1 | 1000 | 5.00E+04 | 100 | 1.44E+02 | 1 | 1000 | 5.00E+04 | 100 | 1.46E+02 |
| 0.1 | 1000 | 1.00E+05 | 100 | 3.03E+02 | 1 | 1000 | 1.00E+05 | 100 | 3.03E+02 |
| 0.1 | 1000 | 2.00E+05 | 100 | 5.93E+01 | 1 | 1000 | 2.00E+05 | 100 | 5.95E+01 |
| 0.1 | 1000 | 5.00E+05 | 100 | 1.53E+03 | 1 | 1000 | 5.00E+05 | 100 | 1.53E+03 |
| 0.1 | 1000 | 1.00E+06 | 100 | 3.11E+03 | 1 | 1000 | 1.00E+06 | 100 | 3.04E+03 |

Table 4: CAKS test results for the cases $\sigma^2 \in \{0.8^2, 0.9^2, 0.95^2\}$ of scenario S2.

| $\sigma$ | $T$ | $J$ | Rejections | Avg. Time | $\sigma$ | $T$ | $J$ | Rejections | Avg. Time | $\sigma$ | $T$ | $J$ | Rejections | Avg. Time |
|---|---|---|---|---|---|---|---|---|---|---|---|---|---|---|
| 0.8 | 100 | 1.00E+02 | 100 | 4.84E-03 | 0.9 | 100 | 1.00E+02 | 95 | 4.81E-03 | 0.95 | 100 | 1.00E+02 | 29 | 4.83E-03 |
| 0.8 | 100 | 2.00E+02 | 100 | 7.02E-03 | 0.9 | 100 | 2.00E+02 | 100 | 6.83E-03 | 0.95 | 100 | 2.00E+02 | 51 | 6.96E-03 |
| 0.8 | 100 | 5.00E+02 | 100 | 1.38E-02 | 0.9 | 100 | 5.00E+02 | 100 | 1.34E-02 | 0.95 | 100 | 5.00E+02 | 95 | 1.37E-02 |
| 0.8 | 100 | 1.00E+03 | 100 | 2.52E-02 | 0.9 | 100 | 1.00E+03 | 100 | 2.47E-02 | 0.95 | 100 | 1.00E+03 | 100 | 2.49E-02 |
| 0.8 | 100 | 2.00E+03 | 100 | 4.89E-02 | 0.9 | 100 | 2.00E+03 | 100 | 4.81E-02 | 0.95 | 100 | 2.00E+03 | 100 | 4.81E-02 |
| 0.8 | 100 | 5.00E+03 | 100 | 1.22E-01 | 0.9 | 100 | 5.00E+03 | 100 | 1.17E-01 | 0.95 | 100 | 5.00E+03 | 100 | 1.20E-01 |
| 0.8 | 100 | 1.00E+04 | 100 | 2.50E-01 | 0.9 | 100 | 1.00E+04 | 100 | 2.38E-01 | 0.95 | 100 | 1.00E+04 | 100 | 2.43E-01 |
| 0.8 | 100 | 2.00E+04 | 100 | 5.18E-01 | 0.9 | 100 | 2.00E+04 | 100 | 4.88E-01 | 0.95 | 100 | 2.00E+04 | 100 | 4.99E-01 |
| 0.8 | 100 | 5.00E+04 | 100 | 1.38E+00 | 0.9 | 100 | 5.00E+04 | 100 | 1.29E+00 | 0.95 | 100 | 5.00E+04 | 100 | 1.31E+00 |
| 0.8 | 100 | 1.00E+05 | 100 | 2.91E+00 | 0.9 | 100 | 1.00E+05 | 100 | 2.82E+00 | 0.95 | 100 | 1.00E+05 | 100 | 2.86E+00 |
| 0.8 | 100 | 2.00E+05 | 100 | 6.11E+00 | 0.9 | 100 | 2.00E+05 | 100 | 6.06E+00 | 0.95 | 100 | 2.00E+05 | 100 | 6.01E+00 |
| 0.8 | 100 | 5.00E+05 | 100 | 1.45E+01 | 0.9 | 100 | 5.00E+05 | 100 | 1.43E+01 | 0.95 | 100 | 5.00E+05 | 100 | 1.45E+01 |
| 0.8 | 100 | 1.00E+06 | 100 | 2.92E+01 | 0.9 | 100 | 1.00E+06 | 100 | 2.92E+01 | 0.95 | 100 | 1.00E+06 | 100 | 2.96E+01 |
| 0.8 | 1000 | 1.00E+02 | 100 | 4.79E-02 | 0.9 | 1000 | 1.00E+02 | 100 | 4.72E-02 | 0.95 | 1000 | 1.00E+02 | 92 | 4.78E-02 |
| 0.8 | 1000 | 2.00E+02 | 100 | 6.90E-02 | 0.9 | 1000 | 2.00E+02 | 100 | 6.88E-02 | 0.95 | 1000 | 2.00E+02 | 100 | 6.97E-02 |
| 0.8 | 1000 | 5.00E+02 | 100 | 1.35E-01 | 0.9 | 1000 | 5.00E+02 | 100 | 1.36E-01 | 0.95 | 1000 | 5.00E+02 | 100 | 1.37E-01 |
| 0.8 | 1000 | 1.00E+03 | 100 | 2.46E-01 | 0.9 | 1000 | 1.00E+03 | 100 | 2.47E-01 | 0.95 | 1000 | 1.00E+03 | 100 | 2.49E-01 |
| 0.8 | 1000 | 2.00E+03 | 100 | 4.76E-01 | 0.9 | 1000 | 2.00E+03 | 100 | 4.75E-01 | 0.95 | 1000 | 2.00E+03 | 100 | 4.86E-01 |
| 0.8 | 1000 | 5.00E+03 | 100 | 1.20E+00 | 0.9 | 1000 | 5.00E+03 | 100 | 1.19E+00 | 0.95 | 1000 | 5.00E+03 | 100 | 1.22E+00 |
| 0.8 | 1000 | 1.00E+04 | 100 | 2.45E+00 | 0.9 | 1000 | 1.00E+04 | 100 | 2.42E+00 | 0.95 | 1000 | 1.00E+04 | 100 | 2.47E+00 |
| 0.8 | 1000 | 2.00E+04 | 100 | 4.98E+00 | 0.9 | 1000 | 2.00E+04 | 100 | 4.96E+00 | 0.95 | 1000 | 2.00E+04 | 100 | 5.07E+00 |
| 0.8 | 1000 | 5.00E+04 | 100 | 1.33E+01 | 0.9 | 1000 | 5.00E+04 | 100 | 1.33E+01 | 0.95 | 1000 | 5.00E+04 | 100 | 1.36E+01 |
| 0.8 | 1000 | 1.00E+05 | 100 | 2.86E+01 | 0.9 | 1000 | 1.00E+05 | 100 | 2.84E+01 | 0.95 | 1000 | 1.00E+05 | 100 | 2.90E+01 |
| 0.8 | 1000 | 2.00E+05 | 100 | 6.09E+01 | 0.9 | 1000 | 2.00E+05 | 100 | 6.10E+01 | 0.95 | 1000 | 2.00E+05 | 100 | 6.04E+01 |
| 0.8 | 1000 | 5.00E+05 | 100 | 1.43E+02 | 0.9 | 1000 | 5.00E+05 | 100 | 1.44E+02 | 0.95 | 1000 | 5.00E+05 | 100 | 1.45E+02 |
| 0.8 | 1000 | 1.00E+06 | 100 | 2.90E+02 | 0.9 | 1000 | 1.00E+06 | 100 | 2.92E+02 | 0.95 | 1000 | 1.00E+06 | 100 | 2.94E+02 |

Table 5: CAKS test results for the cases $\sigma^2 \in \{1.05^2, 1.1^2, 1.2^2\}$ of scenario S2.

| $\sigma$ | $T$ | $J$ | Rejections | Avg. Time | $\sigma$ | $T$ | $J$ | Rejections | Avg. Time | $\sigma$ | $T$ | $J$ | Rejections | Avg. Time |
|---|---|---|---|---|---|---|---|---|---|---|---|---|---|---|
| 1.05 | 100 | 1.00E+02 | 25 | 5.24E-03 | 1.1 | 100 | 1.00E+02 | 60 | 4.81E-03 | 1.2 | 100 | 1.00E+02 | 100 | 4.96E-03 |
| 1.05 | 100 | 2.00E+02 | 27 | 7.23E-03 | 1.1 | 100 | 2.00E+02 | 99 | 7.18E-03 | 1.2 | 100 | 2.00E+02 | 100 | 7.11E-03 |
| 1.05 | 100 | 5.00E+02 | 85 | 1.42E-02 | 1.1 | 100 | 5.00E+02 | 100 | 1.40E-02 | 1.2 | 100 | 5.00E+02 | 100 | 1.36E-02 |
| 1.05 | 100 | 1.00E+03 | 100 | 2.60E-02 | 1.1 | 100 | 1.00E+03 | 100 | 2.55E-02 | 1.2 | 100 | 1.00E+03 | 100 | 2.47E-02 |
| 1.05 | 100 | 2.00E+03 | 100 | 5.04E-02 | 1.1 | 100 | 2.00E+03 | 100 | 4.94E-02 | 1.2 | 100 | 2.00E+03 | 100 | 4.76E-02 |
| 1.05 | 100 | 5.00E+03 | 100 | 1.26E-01 | 1.1 | 100 | 5.00E+03 | 100 | 1.23E-01 | 1.2 | 100 | 5.00E+03 | 100 | 1.19E-01 |
| 1.05 | 100 | 1.00E+04 | 100 | 2.57E-01 | 1.1 | 100 | 1.00E+04 | 100 | 2.50E-01 | 1.2 | 100 | 1.00E+04 | 100 | 2.43E-01 |
| 1.05 | 100 | 2.00E+04 | 100 | 5.29E-01 | 1.1 | 100 | 2.00E+04 | 100 | 5.11E-01 | 1.2 | 100 | 2.00E+04 | 100 | 5.02E-01 |
| 1.05 | 100 | 5.00E+04 | 100 | 1.41E+00 | 1.1 | 100 | 5.00E+04 | 100 | 1.34E+00 | 1.2 | 100 | 5.00E+04 | 100 | 1.32E+00 |
| 1.05 | 100 | 1.00E+05 | 100 | 2.97E+00 | 1.1 | 100 | 1.00E+05 | 100 | 2.85E+00 | 1.2 | 100 | 1.00E+05 | 100 | 2.94E+00 |
| 1.05 | 100 | 2.00E+05 | 100 | 5.97E+00 | 1.1 | 100 | 2.00E+05 | 100 | 5.94E+00 | 1.2 | 100 | 2.00E+05 | 100 | 5.92E+00 |
| 1.05 | 100 | 5.00E+05 | 100 | 1.50E+01 | 1.1 | 100 | 5.00E+05 | 100 | 1.51E+01 | 1.2 | 100 | 5.00E+05 | 100 | 1.47E+01 |
| 1.05 | 100 | 1.00E+06 | 100 | 3.03E+01 | 1.1 | 100 | 1.00E+06 | 100 | 3.07E+01 | 1.2 | 100 | 1.00E+06 | 100 | 2.96E+01 |
| 1.05 | 1000 | 1.00E+02 | 41 | 4.88E-02 | 1.1 | 1000 | 1.00E+02 | 100 | 5.09E-02 | 1.2 | 1000 | 1.00E+02 | 100 | 4.83E-02 |
| 1.05 | 1000 | 2.00E+02 | 93 | 7.13E-02 | 1.1 | 1000 | 2.00E+02 | 100 | 7.30E-02 | 1.2 | 1000 | 2.00E+02 | 100 | 7.02E-02 |
| 1.05 | 1000 | 5.00E+02 | 100 | 1.39E-01 | 1.1 | 1000 | 5.00E+02 | 100 | 1.41E-01 | 1.2 | 1000 | 5.00E+02 | 100 | 1.35E-01 |
| 1.05 | 1000 | 1.00E+03 | 100 | 2.47E-01 | 1.1 | 1000 | 1.00E+03 | 100 | 2.57E-01 | 1.2 | 1000 | 1.00E+03 | 100 | 2.47E-01 |
| 1.05 | 1000 | 2.00E+03 | 100 | 4.79E-01 | 1.1 | 1000 | 2.00E+03 | 100 | 5.01E-01 | 1.2 | 1000 | 2.00E+03 | 100 | 4.81E-01 |
| 1.05 | 1000 | 5.00E+03 | 100 | 1.21E+00 | 1.1 | 1000 | 5.00E+03 | 100 | 1.27E+00 | 1.2 | 1000 | 5.00E+03 | 100 | 1.21E+00 |
| 1.05 | 1000 | 1.00E+04 | 100 | 2.45E+00 | 1.1 | 1000 | 1.00E+04 | 100 | 2.61E+00 | 1.2 | 1000 | 1.00E+04 | 100 | 2.49E+00 |
| 1.05 | 1000 | 2.00E+04 | 100 | 5.01E+00 | 1.1 | 1000 | 2.00E+04 | 100 | 5.34E+00 | 1.2 | 1000 | 2.00E+04 | 100 | 5.10E+00 |
| 1.05 | 1000 | 5.00E+04 | 100 | 1.37E+01 | 1.1 | 1000 | 5.00E+04 | 100 | 1.44E+01 | 1.2 | 1000 | 5.00E+04 | 100 | 1.38E+01 |
| 1.05 | 1000 | 1.00E+05 | 100 | 2.94E+01 | 1.1 | 1000 | 1.00E+05 | 100 | 3.07E+01 | 1.2 | 1000 | 1.00E+05 | 100 | 2.95E+01 |
| 1.05 | 1000 | 2.00E+05 | 100 | 6.00E+01 | 1.1 | 1000 | 2.00E+05 | 100 | 5.98E+01 | 1.2 | 1000 | 2.00E+05 | 100 | 5.95E+01 |
| 1.05 | 1000 | 5.00E+05 | 100 | 1.51E+02 | 1.1 | 1000 | 5.00E+05 | 100 | 1.50E+02 | 1.2 | 1000 | 5.00E+05 | 100 | 1.46E+02 |
| 1.05 | 1000 | 1.00E+06 | 100 | 3.06E+02 | 1.1 | 1000 | 1.00E+06 | 100 | 3.00E+02 | 1.2 | 1000 | 1.00E+06 | 100 | 2.97E+02 |

Table 6: CAKS test results for the cases df $\in \{1, 5, 10\}$ of scenario S3.

| df | $T$ | $J$ | Rejections | Avg. Time | df | $T$ | $J$ | Rejections | Avg. Time | df | $T$ | $J$ | Rejections | Avg. Time |
|---|---|---|---|---|---|---|---|---|---|---|---|---|---|---|
| 1 | 100 | 1.00E+02 | 100 | 8.64E-03 | 5 | 100 | 1.00E+02 | 63 | 1.01E-02 | 10 | 100 | 1.00E+02 | 19 | 9.46E-03 |
| 1 | 100 | 2.00E+02 | 100 | 1.45E-02 | 5 | 100 | 2.00E+02 | 99 | 1.67E-02 | 10 | 100 | 2.00E+02 | 36 | 1.43E-02 |
| 1 | 100 | 5.00E+02 | 100 | 3.25E-02 | 5 | 100 | 5.00E+02 | 100 | 3.71E-02 | 10 | 100 | 5.00E+02 | 88 | 3.11E-02 |
| 1 | 100 | 1.00E+03 | 100 | 6.30E-02 | 5 | 100 | 1.00E+03 | 100 | 7.15E-02 | 10 | 100 | 1.00E+03 | 100 | 5.98E-02 |
| 1 | 100 | 2.00E+03 | 100 | 1.24E-01 | 5 | 100 | 2.00E+03 | 100 | 1.40E-01 | 10 | 100 | 2.00E+03 | 100 | 1.16E-01 |
| 1 | 100 | 5.00E+03 | 100 | 3.12E-01 | 5 | 100 | 5.00E+03 | 100 | 3.50E-01 | 10 | 100 | 5.00E+03 | 100 | 2.92E-01 |
| 1 | 100 | 1.00E+04 | 100 | 6.29E-01 | 5 | 100 | 1.00E+04 | 100 | 7.05E-01 | 10 | 100 | 1.00E+04 | 100 | 5.86E-01 |
| 1 | 100 | 2.00E+04 | 100 | 1.27E+00 | 5 | 100 | 2.00E+04 | 100 | 1.42E+00 | 10 | 100 | 2.00E+04 | 100 | 1.19E+00 |
| 1 | 100 | 5.00E+04 | 100 | 3.27E+00 | 5 | 100 | 5.00E+04 | 100 | 3.63E+00 | 10 | 100 | 5.00E+04 | 100 | 3.07E+00 |
| 1 | 100 | 1.00E+05 | 100 | 6.70E+00 | 5 | 100 | 1.00E+05 | 100 | 7.51E+00 | 10 | 100 | 1.00E+05 | 100 | 6.12E+00 |
| 1 | 100 | 2.00E+05 | 100 | 1.31E+01 | 5 | 100 | 2.00E+05 | 100 | 1.44E+01 | 10 | 100 | 2.00E+05 | 100 | 1.24E+01 |
| 1 | 100 | 5.00E+05 | 100 | 3.37E+01 | 5 | 100 | 5.00E+05 | 100 | 3.76E+01 | 10 | 100 | 5.00E+05 | 100 | 3.12E+01 |
| 1 | 100 | 1.00E+06 | 100 | 6.76E+01 | 5 | 100 | 1.00E+06 | 100 | 7.58E+01 | 10 | 100 | 1.00E+06 | 100 | 6.27E+01 |
| 1 | 1000 | 1.00E+02 | 100 | 8.75E-02 | 5 | 1000 | 1.00E+02 | 100 | 9.81E-02 | 10 | 1000 | 1.00E+02 | 38 | 8.51E-02 |
| 1 | 1000 | 2.00E+02 | 100 | 1.47E-01 | 5 | 1000 | 2.00E+02 | 100 | 1.66E-01 | 10 | 1000 | 2.00E+02 | 88 | 1.41E-01 |
| 1 | 1000 | 5.00E+02 | 100 | 3.29E-01 | 5 | 1000 | 5.00E+02 | 100 | 3.70E-01 | 10 | 1000 | 5.00E+02 | 100 | 3.06E-01 |
| 1 | 1000 | 1.00E+03 | 100 | 6.33E-01 | 5 | 1000 | 1.00E+03 | 100 | 7.13E-01 | 10 | 1000 | 1.00E+03 | 100 | 5.92E-01 |
| 1 | 1000 | 2.00E+03 | 100 | 1.25E+00 | 5 | 1000 | 2.00E+03 | 100 | 1.41E+00 | 10 | 1000 | 2.00E+03 | 100 | 1.16E+00 |
| 1 | 1000 | 5.00E+03 | 100 | 3.13E+00 | 5 | 1000 | 5.00E+03 | 100 | 3.52E+00 | 10 | 1000 | 5.00E+03 | 100 | 2.90E+00 |
| 1 | 1000 | 1.00E+04 | 100 | 6.29E+00 | 5 | 1000 | 1.00E+04 | 100 | 7.10E+00 | 10 | 1000 | 1.00E+04 | 100 | 5.84E+00 |
| 1 | 1000 | 2.00E+04 | 100 | 1.27E+01 | 5 | 1000 | 2.00E+04 | 100 | 1.43E+01 | 10 | 1000 | 2.00E+04 | 100 | 1.18E+01 |
| 1 | 1000 | 5.00E+04 | 100 | 3.27E+01 | 5 | 1000 | 5.00E+04 | 100 | 3.68E+01 | 10 | 1000 | 5.00E+04 | 100 | 3.04E+01 |
| 1 | 1000 | 1.00E+05 | 100 | 6.70E+01 | 5 | 1000 | 1.00E+05 | 100 | 7.50E+01 | 10 | 1000 | 1.00E+05 | 100 | 6.22E+01 |
| 1 | 1000 | 2.00E+05 | 100 | 1.32E+02 | 5 | 1000 | 2.00E+05 | 100 | 1.45E+02 | 10 | 1000 | 2.00E+05 | 100 | 1.24E+02 |
| 1 | 1000 | 5.00E+05 | 100 | 3.40E+02 | 5 | 1000 | 5.00E+05 | 100 | 3.71E+02 | 10 | 1000 | 5.00E+05 | 100 | 3.21E+02 |
| 1 | 1000 | 1.00E+06 | 100 | 6.94E+02 | 5 | 1000 | 1.00E+06 | 100 | 7.58E+02 | 10 | 1000 | 1.00E+06 | 100 | 6.51E+02 |

Table 7: CAKS test results for the cases df $\in \{50, 100\}$ of scenario S3.

| df | $T$ | $J$ | Rejections | Avg. Time | df | $T$ | $J$ | Rejections | Avg. Time |
|---|---|---|---|---|---|---|---|---|---|
| 50 | 100 | 1.00E+02 | 14 | 8.26E-03 | 100 | 100 | 1.00E+02 | 7 | 8.25E-03 |
| 50 | 100 | 2.00E+02 | 11 | 1.31E-02 | 100 | 100 | 2.00E+02 | 13 | 1.32E-02 |
| 50 | 100 | 5.00E+02 | 11 | 2.86E-02 | 100 | 100 | 5.00E+02 | 10 | 2.85E-02 |
| 50 | 100 | 1.00E+03 | 13 | 5.44E-02 | 100 | 100 | 1.00E+03 | 13 | 5.42E-02 |
| 50 | 100 | 2.00E+03 | 19 | 1.07E-01 | 100 | 100 | 2.00E+03 | 13 | 1.07E-01 |
| 50 | 100 | 5.00E+03 | 52 | 2.61E-01 | 100 | 100 | 5.00E+03 | 11 | 2.69E-01 |
| 50 | 100 | 1.00E+04 | 79 | 5.33E-01 | 100 | 100 | 1.00E+04 | 23 | 5.40E-01 |
| 50 | 100 | 2.00E+04 | 100 | 1.06E+00 | 100 | 100 | 2.00E+04 | 43 | 1.08E+00 |
| 50 | 100 | 5.00E+04 | 100 | 2.78E+00 | 100 | 100 | 5.00E+04 | 94 | 2.76E+00 |
| 50 | 100 | 1.00E+05 | 100 | 5.76E+00 | 100 | 100 | 1.00E+05 | 100 | 5.69E+00 |
| 50 | 100 | 2.00E+05 | 100 | 1.45E+01 | 100 | 100 | 2.00E+05 | 100 | 1.35E+01 |
| 50 | 100 | 5.00E+05 | 100 | 2.95E+01 | 100 | 100 | 5.00E+05 | 100 | 2.93E+01 |
| 50 | 100 | 1.00E+06 | 100 | 5.95E+01 | 100 | 100 | 1.00E+06 | 100 | 5.91E+01 |
| 50 | 1000 | 1.00E+02 | 6 | 8.07E-02 | 100 | 1000 | 1.00E+02 | 11 | 8.01E-02 |
| 50 | 1000 | 2.00E+02 | 11 | 1.33E-01 | 100 | 1000 | 2.00E+02 | 5 | 1.31E-01 |
| 50 | 1000 | 5.00E+02 | 14 | 2.89E-01 | 100 | 1000 | 5.00E+02 | 14 | 2.84E-01 |
| 50 | 1000 | 1.00E+03 | 17 | 5.52E-01 | 100 | 1000 | 1.00E+03 | 8 | 5.43E-01 |
| 50 | 1000 | 2.00E+03 | 46 | 1.08E+00 | 100 | 1000 | 2.00E+03 | 26 | 1.08E+00 |
| 50 | 1000 | 5.00E+03 | 99 | 2.71E+00 | 100 | 1000 | 5.00E+03 | 19 | 2.74E+00 |
| 50 | 1000 | 1.00E+04 | 100 | 5.47E+00 | 100 | 1000 | 1.00E+04 | 61 | 5.46E+00 |
| 50 | 1000 | 2.00E+04 | 100 | 1.11E+01 | 100 | 1000 | 2.00E+04 | 100 | 1.11E+01 |
| 50 | 1000 | 5.00E+04 | 100 | 2.88E+01 | 100 | 1000 | 5.00E+04 | 100 | 2.87E+01 |
| 50 | 1000 | 1.00E+05 | 100 | 5.90E+01 | 100 | 1000 | 1.00E+05 | 100 | 6.08E+01 |
| 50 | 1000 | 2.00E+05 | 100 | 1.45E+02 | 100 | 1000 | 2.00E+05 | 100 | 1.35E+02 |
| 50 | 1000 | 5.00E+05 | 100 | 2.98E+02 | 100 | 1000 | 5.00E+05 | 100 | 3.02E+02 |
| 50 | 1000 | 1.00E+06 | 100 | 6.08E+02 | 100 | 1000 | 1.00E+06 | 100 | 5.93E+02 |

indicates that the CAKS test is behaving as expected when the null hypothesis is true. That is, the number of rejections are all approximately 10 or there about, as expected. We note that the 95% Wald margin of error for the number of rejections (out of 100) under the null at the $\alpha = 0.1$ level is $\pm 6$, which corresponds well with our observations.

Next, as a general note, we observe throughout all of the tables that the CAKS test becomes more powerful as $N$ increases, both via an increase in $T$ and $J$. This corresponds well with intuition regarding large sample behaviors of test, as well as with the conclusions of Propositions 2 and 3. Another general note is that the average time taken to compute the CAKS statistics are increasing in $N$, both via increases in $T$ and $J$. Although not much can be said regarding the effect of $T$ on computation time, as we only simulate two values of $T$, we can observe that the computation time appears to be linearly increasing in $J$. Since the KS statistic requires an ordering to compute, this result corresponds well with the fact that most sorting algorithms have $O(N)$ average case time complexity (cf. Cormen et al., 2002, Ch. 8).

Tables 2–7 indicate that it is possible to improve the power of the CAKS test in order to assess whether data arises from a distribution with a differing means (S1), variances (S2), or shape (S3), to that which is hypothesized. The results from the tables also show that the CAKS test becomes more powerful in making these distinctions as more observations, $N$, are made available. The numerical study has therefore successfully achieved the goal of demonstrating that the CAKS test behaves as is expected of a hypothesis testing procedure, and as is expected from the conclusions of Propositions 2 and 3.

We finally consider the computation time of the CAKS test in comparison to the KS test, conducted on the same number of observations. Performing a KS test of the same null and simulation distribution as the $\mu = 0$ case of S1, we obtain the average computation times (from 100 repetitions) of 3.33E-03, 2.29E-02, 2.51E-01, 3.09E+00, 4.31E+01, 1.01E+04, for $N \in 10^{\{4,5,6,7,8,9\}}$, respectively. The KS tests were conducted via the ks.test() function.

Comparing the results above with those from the rows of Tables 2–5 with equal values of $N$, we observe that the CAKS test is slightly slower than the batch KS test for $N = 10^4$, and approximately equal in computation time for $N \in 10^{\{5,6,7\}}$. The CAKS test then becomes slightly faster for $N = 10^8$, averaging approximately 3/4 the computation time of the KS test. When $N = 10^9$, the CAKS test is two orders of magnitude faster than the KS test. We note that there are no boundaries to increasing the the application of the CAKS test beyond the assessed values of $N$, and that in practice, one computes the CAKS statistic in an online manner, which implies that the computation time should be considered incrementally at each time period rather than all together, as we have done. We do not make comparisons to the timing results from Tables 6 and 7, as the computation of the $t$ distribution function is slower than the computation of $\Phi$ in R, as can be observed from the direct comparisons in respective cases between Tables 2–5, and 6 and 7.

We found it difficult to make comparisons with the batch KS test for larger values of $N$, as the computations required an infeasible amount of time. We believe that the evaluation of the asymptotic probability of the test statistic to a reasonable degree of accuracy takes up the majority of the time needed for conducting the KS test via the ks.test() function.

## 5. Conclusions

The Big Data setting prevents the use of numerous useful tools that do not scale well with increasing numbers of observations. One such tool is the KS test for the GoF of data to some known distribution. Furthermore, the KS test is also inappropriate in the Big Data setting as data is often obtained as a strream, rather than in batch.

We have presented an alternative to the KS test, which scales linearly, on average, in computation time and can be applied to streamed data. Our alternative testing procedure is based on the CA estimators of Li et al. (2013) and Matloff (2016b), and is thus called the CAKS test for GoF.

Using results from Li et al. (2013), we proved that the CAKS test statistic is asymptotically normal, under the null hypothesis that data arises from some known distribution. Further theoretical proofs demonstrated that the CAKS test is both consistent when the alternative hypothesis is fixed and when it is approaching the null at a sufficiently slow rate. These consistency results mirror those that can be obtained for the usual batch KS test.

In addition to our test derivation and theoretical proofs, we also conducted a numerical study. In our numerical study, we found that the CAKS test behaved as expected when the sample size $N$ increased, either through increases of the chunk size $J$ or number of chunks $T$. That is, we found that the test became more powerful, under all scenarios as $N$ increased. Furthermore, we found that the test was capable of differentiate between the null alternative distributions with differing means, variances, and shapes, in three simulation scenarios.

Using the R programming language and environment, we found the the CAKS test was faster than the KS test for large $N$, especially when $N = 10^9$, where the CAKS test was two orders of magnitude faster in computational time. This makes it a far more appropriate method for GoF testing than the KS test in the Big Data setting.

Lastly, we note that the techniques for test construction that we have developed in this paper are fully transferrable to the construction of tests that are based on other statistics with nonstandard finite sample and asymptotic distributions. For example, one could adapt the methods from this paper to construct a Big Data and stream-suitable Anderson-Darling test (Anderson and Darling, 1954) or Lilliefors test (Lilliefors, 1967).

## References

Abad, C. L., Luu, H., Roberts, N., Lee, K., Lu, Y., Campbell, R. H., 2012. Metadata traces and workload models for evaluating big storage systems. In: Proceedings of the IEEE/ACM Fifth International Conference on Utility and Cloud Computing. pp. 125–135.

Anderson, T. W., Darling, D. A., 1954. A test for Goodness-of-fit. Journal of the American Statistical Association 49, 765–769.

Bifet, A., Holmes, G., Kirkby, R., Pfahringer, B., 2010. MOA: massive online analysis. Journal of Machine Learning Research 11, 1601–1604.

Buoncristiano, M., Mecca, G., Quitarelli, E., Roveri, M., Santoro, D., Tanca, L., 2015. Database challenges for exploratory computing. ACM SIGMOD Record 44, 17–22.

Chen, M., Mao, S., Liu, Y., 2014. Big Data: a survey. Mobile Networks and Applications 19, 171–209.

Clark, N. R., Ma'ayan, A., 2011. Introduction to statistical methods for analyzing large data sets: gene-set enrichment analysis. Science Signaling 4, tr4.

Cormen, T. H., Leiserson, C. E., Rivest, R. L., Stein, C., 2002. Introduction To Algorithms. MIT Press, Cambridge.

DasGupta, A., 2011. Probability for Statistics and Machine Learning. Springer, New York.

Demirhan, H., Bitirim, N., 2016. CryptRndTest: An R package for testing the cryptographic randomness. The R Journal 8, 233–247.

Drummond, A. J., Ho, S. Y. W., Phillips, M. J., Rambaut, A., 2006. Relaxed phyologenetics and dating with confidence. PloS Biology 4, e88.

Durbin, J., 1972. Distribution Theory for Tests Based on the Sample Distribution Function. SIAM, Philadelphia.

Gruet, M. A., Philippe, A., Robert, C. P., 1998. Discretization and MCMC Converence Assessment. Springer, New York, Ch. Estimation of exponential mixtures, pp. 161–173.

Jacobs, A., 2009. The pathologies of Big Data. Communications of the ACM 52, 36–44.

Kolmogorov, A., 1933. Sulla determinazione empirica di una legge di distributione. Giornale dell' Istituto Italiano degli Attuari 4, 83–91.

Lall, A., 2015. Data streaming algorithm for the Kolmogorov-Smirnov test. In: Proceedings of the IEEE Iinternational Conference on Big Data. pp. 95–104.

Lehmann, E. L., Romano, J. P., 2005. Testing Statistical Hypotheses. Springer, New York.

Li, R., Lin, D. K. J., Li, B., 2013. Statistical inference in massive data sets. Applied Stochastic Models in Business and Industry 29, 399–409.

Lilliefors, H., 1967. On the Kolmogorov-Smirnov test for normality with mean and variance unknown. Journal of the American Statistical Association 62, 399–402.

Macqueen, D. J., Fuentes, E. N., Valdes, J. A., Molina, A., Martin, S. A. M., 2014. The vertebrate muscle-specific RING finger protein family includes

MuRF4 – A novel, conserved E3-ubiquitin ligase. FEBS Letters 588, 4390–4397.

Matloff, N., 2016a. Parallel Computing for Data Science: With Examples in R, C++ and CUDA. CRC Press, Boca Raton.

Matloff, N., 2016b. Software alchemy: Turning complex statistical computations into embarrassingly-parallel ones. Journal of Statistical Software 71, 1–15.

Nguyen, C. D., Carlin, J. B., Lee, K. J., 2013. Diagnosing problems with imputation models using the Kolmogorov-Smirnov test: a simulation study. BMC Medical Research Methodology 13, 144.

Nguyen, H. D., 2017. A simple online parameter estimation technique with asymptotic guarantees. ArXiv:1703.07039.

Nguyen, H. D., McLachlan, G. J., 2017. Chunked-and-averaged estimators for vector parameters. ArXiv:1612.06492.

R Core Team, 2016. R: a language and environment for statistical computing. R Foundation for Statistical Computing.

Robert, C. P., Casella, G., 2010. Introducing Monte Carlo Methods with R. Springer, New York.

Smirnov, N., 1948. Table for estimating the goodness of fit of empirical distributions. Annals of Mathematical Statistics 19, 297–281.

Tran, D.-H., Gaber, M. M., Sattler, K.-U., 2014. Change detection in streaming data in the era of Big Data: models and issues. ACM SIGKDD Explorations 16, 30–38.

Vermeesch, P., Garzaanti, E., 2015. Making geological sense of 'Big Data' in sedimentary provenance analysis. Chemical Geology 409, 20–27.

Wang, J., Tsang, W. W., Marsaglia, G., 2003. Evaluating Kolmogorov's distribution. Journal of Statistical Software 8, 18.

Wang, L., Luo, G., Yi, K., Cormode, G., 2013. Quantiles over data streams: an experimental study. In: Proceedings of SIGMOD. pp. 1–12.